%% file: Sage_LaTeX_Guidelines.tex
\documentclass[Afour, times, sagev]{sagej}

\usepackage{moreverb,url}

\usepackage[colorlinks,bookmarksopen,bookmarksnumbered,citecolor=red,urlcolor=red]{hyperref}
\usepackage{enumitem}
\usepackage{soul}
\usepackage{amsmath}
\usepackage{amssymb}
\usepackage{xspace}
\usepackage{multirow}
\usepackage[table]{xcolor}
\usepackage[svgnames, dvipsnames]{xcolor}
\usepackage{multicol}
\usepackage{xcolor}
\usepackage{multibib}
\usepackage{cleveref}
\usepackage{fancyvrb}
\usepackage{tikz}
\usepackage{booktabs}
\usepackage{ulem}

\newcommand\BibTeX{{\rmfamily B\kern-.05em \textsc{i\kern-.025em b}\kern-.08em
T\kern-.1667em\lower.7ex\hbox{E}\kern-.125emX}}

\newcommand{\user}{Ryan\xspace}
\newcommand{\sys}{\textcolor{black}{\mbox{\textsc{Asap}}}\xspace}
\newcites{appx}{References}

\definecolor{cback}{HTML}{E9ECEF}
\definecolor{cframe}{HTML}{495057}
\definecolor{clightblue}{HTML}{DDEFFC}
\definecolor{clightgreen}{HTML}{DFFFE5}
\definecolor{clightyellow}{HTML}{FFF4DB}

\definecolor{cback1}{HTML}{F8F9FA} 
\definecolor{cframe1}{HTML}{ADB5BD} 

\definecolor{cback2}{HTML}{F1F3F5} 
\definecolor{cframe2}{HTML}{CED4DA} 

\definecolor{cback3}{HTML}{EBF5FB} 
\definecolor{cframe3}{HTML}{A9CCE3} 

\definecolor{cback4}{HTML}{FDF3E7} 
\definecolor{cframe4}{HTML}{F5CBA7} 

\newcommand*\circled[1]{\tikz[baseline=(char.base)]{
    \node[shape=rectangle,rounded corners=1.5pt,fill=cback,text=black,draw=cframe,inner sep=1pt] (char) {#1};}}

\newcommand*\circledVarD[1]{\tikz[baseline=(char.base)]{
    \node[shape=rectangle,rounded corners=1.5pt,fill=cback4,text=black,draw=cframe4,inner sep=1pt] (char) {#1};}}

\def\volumeyear{2016}

\newcommand{\revsec}[1]{{\color{black}#1}}
\newcommand{\rev}[1]{{\color{black}#1}}

\DeclareRobustCommand{\erase}[1]{}

\begin{document}


\title{ASAP: Visual Analytics for Identifying and Analyzing Image Patterns in AI-generated Images \erase{at Scale}}

\author{Jinbin Huang\affilnum{1}, Yuki Ueno\affilnum{1}, Chen Chen\affilnum{2}, Aditi Mishra\affilnum{3}, Bum Chul Kwon\affilnum{4}, Zhicheng Liu\affilnum{2}, and Chris Bryan\affilnum{1}}

\affiliation{\affilnum{1}Arizona State University, AZ, US\\
\affilnum{2}University of Maryland, MD, US\\
\affilnum{3}Fujitsu Research of America, PA, US\\
\affilnum{4}IBM Research, MA, US
}

\email{yueno@asu.edu}

\corrauth{Yuki Ueno, Arizona State University
Sonoran Visualization Laboratory,
699 S Mill Ave,
Tempe, AZ, US.}

\begin{abstract}
Generative image models can produce highly realistic images, raising concerns about potential misuse in creating deceptive content. Current deepfake approaches face several challenges, including limited generalizability, lack of interpretability, and poor \rev{actionability}. 
To help address these, we present \sys, an interactive visualization system designed to empower users in the analysis and summarization of deceptive patterns in AI-generated images. 
\sys introduces a novel CLIP-adapted image encoder that generates interpretable representations, enabling the \erase{precise} extraction of influential pixel regions via calculated masks. This approach facilitates the identification of key deceptive features through influence measurement techniques. 
These backend techniques are integrated into a visual analytics dashboard that 
allows users to quantify and analyze authenticity-indicative patterns \rev{in image collections containing both authentic and AI-generated images}. This approach also supports the comparative analysis of various generative models, including GANs and diffusion models. 
We demonstrate \sys's efficacy through a user study and two application scenarios using established fake image detection benchmarks, showcasing its ability to effectively extract and quantify deceptive patterns.
\end{abstract}


\maketitle
\input{documents/1-introduction}
\input{documents/2-related_works}
\input{documents/3-design_challenges_and_goals}
\input{documents/4-backend}
\input{documents/5-interface}
\input{documents/6-usage-scenario}
\input{documents/7-user-study}

\input{documents/8-discussion_and_future_work}


\bibliographystyle{SageV}
\bibliography{references_pruned}

\clearpage

\input{documents/appendix}
\bibliographystyleappx{SageV}
\bibliographyappx{references_pruned}

\end{document}

%% file: documents/1-introduction.tex
\section{Introduction}
\label{sec:intro}

\begin{figure*}[t]
  \includegraphics[width=.8\linewidth]{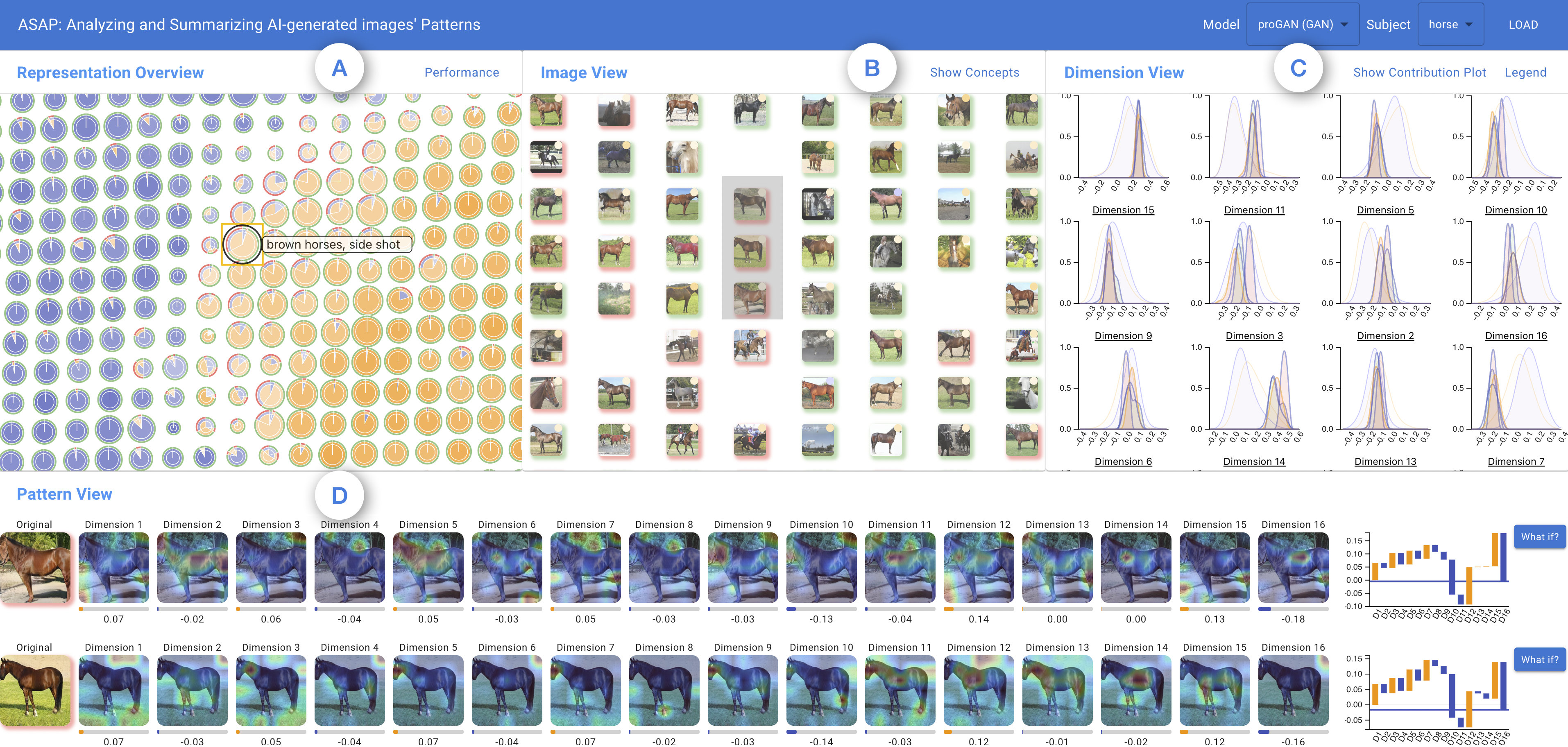}
  \centering
  \caption{
    \sys supports interactively analyzing deceptive patterns in AI-generated images that fool trained classifiers. In this example, a user is analyzing ``brown horses, side shot'' images generated by proGAN.
    \textbf{(A)} The Representation Overview projects image representations into 2D space, grouping them into cells with novel glyphs that summarize classification performance. 
    \textbf{(B)} The Image View highlights fake images misclassified as authentic within selected cells. \textbf{(C)} The Dimension View visualizes embedding distributions across CLIP dimensions, correlating visual patterns with their quantitative characteristics. \textbf{(D)} The Pattern View reveals the most misleading pixel groups through saliency maps, uncovering distinct patterns in AI-generated images.
    }
  \label{fig:teaser}
\end{figure*}

Given the rapid advancement of generative AI (GenAI), image models such as DALL·E 3~\cite{dalle3} and Midjourney~\cite{midjourney} can now create high-fidelity, realistic images that easily deceive humans~\cite{cao2023comprehensive}. This paradigm shift is coinciding with rising demands for effective detection and analysis of realistic-but-fake images~\cite{hamed2024safeguarding}, 
as their misuse can have widespread impacts including misinformation spread~\cite{combating2023Xu}, copyright violation~\cite{sag2023copyright}, 
and credit–blame asymmetry~\cite{porsdam2023generative}. 

While the research community has developed examples of automated detection methods for fake images (that can outperform human experts under certain conditions~\cite{ha2024organic}), these tools often face significant challenges, including \textit{generalizability}~\cite{zhang2019detecting}, \textit{interpretability}~\cite{lloyd2023there, weber2023testing, sadasivan2023can}, and \textit{\rev{actionability}}~\cite{lloyd2023there}. Generalizability challenges arise, in part, from distinct visual signatures across different generative models (e.g., GANs vs. diffusion models \cite{corvi2023detection}), where detectors effective for one model can fail for others~\cite{zhang2019detecting}. Interpretability and \rev{actionability} challenges can arise because black-box detector models output only predictions and do not provide transparency or explanations of decision-making processes~\cite{lloyd2023there}.

\rev{Making these challenges even more salient to real-world use, the types of users and stakeholders who are performing deepfake analysis is also broadening. Such users now include not only ``AI experts'' (such as GenAI developers and researchers), but also ``model users''~\cite{hohman2018visual} such as misinformation researchers, digital forensic analysts, social media moderators, computational journalists, and legal experts, who must analyze and explain deepfake images as part of their workflows~\cite{sohrawardi2024dungeons}. These users often have high computational proficiency on tasks such as pixel-based review and analysis (e.g., via photo editing tools), but lack formal training in the technical and mathematical aspects of AI/ML engineering and prediction. However, they also need to understand deceptive patterns in AI-generated images, including why an image might appear real or fake to a detector.}

Along these lines, there is increasing recognition we need novel human-in-the-loop (HITL) workflows to help 
in analyzing and understanding AI-generated images and GenAI models~\cite{ha2024organic}. Such advocacy has highlighted a need to create interpretable tools and techniques to help overcome the above-described challenges~\cite{weber2023testing}. 

This paper investigates how to address these needs via a design study process~\cite{sedlmair2012design}. Specifically, we develop \sys (Analyzing and Summarizing AI-generated image Patterns), a HITL pipeline and visual analytics tool to help identify, analyze, and summarize patterns in \rev{images ensembles containing a combination of both authentic and fake images}. \sys is designed for analysis settings where image authenticity labels are available, such as curated research datasets, benchmarking environments, or forensic investigations. It includes an end-to-end pipeline for efficient detection and analysis of image patterns in datasets that contain a combination of real and AI-generated images, including an interpretability-focused detector that produces compact embeddings with latent authenticity indicators to facilitate the discovery and explanation of how such patterns can differ among authentic and synthetic images (\rev{i.e., helping users reason about why an image appears real or fake}). \sys's frontend is a visual analytics interface composed of several coordinated views supporting such analysis and exploration, from broad overviews to fine-grained details.

We robustly \rev{evaluate} \sys via a set of usage scenarios and a mixed methods user study. Our results demonstrate how \sys can effectively support analyzing patterns in AI-generated and real images \rev{in an interpretable manner}. Based on the process of designing, implementing, and validating \sys, we also reflect on a number of takeaways and lessons learned, such as how HITL and visualization-driven approaches can be tailored to effectively support generalizability, interpretability, and \rev{actionability} in the context of deepfake analysis, and how tools like \sys address fast-evolving research questions in GenAI trust and authenticity.

%% file: documents/2-related_works.tex
\section{Related Work}
\label{sec:relatedwork}

\sys builds upon prior research in the detection of AI-generated images, explanations for transformer models, and applications of visual analytics for AI and GenAI.

\subsection{Detecting Fake Images}

Fake image detection predates generative image models, as traditional fake image creation and manipulation methods (e.g., via software like Photoshop) were known to introduce detectable changes in images, including visual cues like compression artifacts \cite{agarwal2017photo}, resampling~\cite{popescu2005exposing}, and unnatural reflections~\cite{o2012exposing}. However, many early detectors and classifiers faced generalizability challenges, such as not handling out-of-distribution data or newer models~\cite{cozzolino2018forensictransfer, zhang2019detecting}, prompting the development of more versatile classifiers~\cite{chai2020makes, asnani2022proactive, Ojha_2023_CVPR, zhu2023gendet}.

A significant challenge for learning-based methods is finding a robust feature space to distinguish between authentic and AI-generated images across various models~\cite{zhu2023gendet}. Recent research suggests using feature spaces from large pre-trained models like CLIP~\cite{Ojha_2023_CVPR}, which have shown to be effective and generalizable~\cite{cozzolino2023raising}. 
In \sys{}, we build upon these findings by modifying CLIP's visual encoder to develop a linear classifier that efficiently encodes images into interpretable representations, facilitating the discovery of distinctive patterns in AI-generated images.


Specifically, CLIP (Contrastive Language-Image Pre-training)~\cite{clip} has emerged as a powerful pre-training paradigm that combines textual supervision with visual cues, and has demonstrated remarkable adaptability and success across a range of semantic tasks including image segmentation\cite{li2022language, rao2022denseclip, luo2023segclip}, object detection~\cite{bangalath2022bridging}, classification~\cite{zhou2022conditional}, and medical imaging~\cite{zhao2023clip}. However, CLIP's entanglement of text and vision poses challenges for predominantly visual tasks. Recent studies have highlighted that CLIP's visual encoder may inadvertently embed textual information into its outputs~\cite{materzynska2022disentangling}. Such unintended textual embeddings can compromise the efficacy of gradient-based methods aimed at identifying crucial non-semantic pixels, consequently affecting the accurate identification of inauthentic image regions.

To address this in \sys, we integrate a ``forget-to-spell'' model: an orthogonal linear projection designed to minimize textual content in CLIP's latent space. This approach removes unwarranted textual embeddings, yielding a predominantly visual-centric representation that better serves our task of fake image detection and pattern discovery.


\subsection{Measuring Token Influence In Transformer Models} 

As transformer models become more popular, their explanations gain importance \cite{barkan2021grad, chefer2021generic}. A key interpretability method for transformers is to attribute influence scores to tokens \cite{chefer2021generic, barkan2021grad}, with terms such as \textit{relevance}, \textit{importance}, and \textit{influence} often used interchangeably~\cite{yuan2021explaining}.



For vision transformers, computing token relevance identifies influential pixels in model decisions~\cite{barkan2021grad}; this is also crucial for detecting AI-generated images. Several methods exist for explaining predictions in transformer-based architectures~\cite{abnar2020quantifying, chefer2021transformer, voita2019analyzing}; we employ Chefer et al.'s~\cite{chefer2021generic} approach due to its generalizability and strong performance. Specifically, we adapt their relevance map technique for our CLIP-based image encoder, with a significant modification: instead of using final logits, we utilize embedding dimensions for relevance computation (see `Backend' section). This approach offers a more fine-grained analysis than the single map generated by binary classification logits.

\subsection{Visual Analytics for AI and Generative AI}

Visual analytics tools have been developed for a variety of HITL workflows for AI and GenAI, including data augmentation~\cite{zhao2021human}, labeling~\cite{hoque2022visual},
exploration and analysis~\cite{gou2020vatld}, 
bias inspection and mitigation~\cite{xie2021fairrankvis}, 
model explanation~\cite{huang2022conceptexplainer}, 
model refinement~\cite{wongsuphasawat2017visualizing}, 
\rev{robustness testing~\cite{li2024robustmap}, and trustworthiness promotion~\cite{chatzimparmpas2025visual}}.
\rev{Alongside these systems, a variety of explainable AI (XAI) methods have been developed to explain model decision-making processes, including through local surrogate models~\cite{ribeiro2016should}, feature attribution~\cite{lundberg2017unified,sundararajan2017axiomatic}, class-discriminative localization~\cite{selvaraju2017grad}, or concept-level analysis~\cite{kim2018interpretability}.
These techniques are commonly applicable to binary classification settings (including real/fake image detection), by showing which input features, pixels, or concepts support a predicted class.}

Specifically for GenAI, rapid advancements in this space have led to increased emphases on developing visual analytics tools specifically for generative models. Recent work has, for example, focused on visualizing transformer attention mechanisms~\cite{yeh2023attentionviz}, 
illustrating diffusion processes~\cite{lee2023diffusion}, and enhancing prompt engineering~\cite{feng2023promptmagician}. Broadly, such tools can aim to unveil the inner workings 
of generative models (particularly those based on 
transformer or diffusion methodologies) or support 
interactions with these models.

\rev{One similar tool to \sys is by Ye et al.~\cite{ye2025unified}, who develop a system for reviewing and comparing human and AI-generated paintings. While they likewise employ a CLIP-based approach, they also design and integrate a number of manual painting-focused aesthetic features (such as composition and use of color), and do not focus on differentiating nuanced pixel-specific segments and patterns that distinguish real-vs.-fake images.
Put another way, \sys{} targets a different analytic gap: using interactive visualization to analyze authentic and synthetic image collections beyond simple fake/real labeling, by tightly linking collection-level summaries of images to the specific image evidences (i.e., pixel patterns) responsible for detector decisions, by integrating an explainability-focused detector with coordinated views that connect dimension-specific relevance maps, detector outcomes, contribution scores, and image collections.
This design helps analysts move from simply observing that a model or collection differs, to understanding why an image succeeds or fails detection, by identifying recurring deceptive patterns and quantifying their contribution across diverse generative models.}

%% file: documents/3-design_challenges_and_goals.tex

\section{Design Challenges and Goals}
\label{sec:design_challenges_and_goals}

Our overarching goal is to investigate how to help users analyze collections of (potentially) fake images, such as understanding how they might contain various deceptive or authenticity-indicative patterns imbued by a creating generative image model, and if these pixel patterns fool a deepfake detector. Based on a meta-analysis of papers (i.e., form the `Related Work' section) and similar HITL AI explainability work (e.g., \rev{Xu et al.}~\cite{xu2023transitioning}),
\rev{we identified several analytic tasks such users 
could perform, including exploring an ensemble of synthetic and fake images to understand their pixel-based patterns and relationships, locating suspicious or anomalous data items, identifying and analyzing groups of items (or individual instances) in the ensemble where a detector (mis)classifies, and understanding where and why such misclassifications happens. In such contexts, users would work with data that includes 2D, pixel-based images, ground-truth labeling as to their authentic or deepfake/synthetic status, and would also need transparent outputs from a deepfake detector (i.e., not just predicting if the image was authentic or not, but doing so in a way that supported explanation or interpretation).}

Unfortunately, there are two high-level challenges that make this a non-trivial process: (i)~we need a method to broadly extract and identify patterns that distinguish fake from real images, both qualitatively and quantitatively; (ii)~we need a user experience that facilitates intuitive, interpretable, and in-depth pattern analysis (e.g., at varying levels of detail). We first formalize these into a set of four specific design challenges: \circled{C1}--\circled{C4}.

\subsection{Design Challenges}
\textbf{\circled{C1} Automatic Discovery of Patterns Present in Fake Images}: In scenarios where classifiers are trained to differentiate between real and AI-generated images, they inherently identify unique patterns of both types \cite{cozzolino2023raising}. The challenge is to devise a way to utilize these classifiers to efficiently extract these distinct patterns amidst irrelevant visual information that might confound the analysis. 

\textbf{\circled{C2} Quantifying Influence of Fake Patterns}: Generative models produce diverse artificial patterns, even within the same model~\cite{corvi2023detection}. Analyzing these patterns requires distinctively identifying and quantitatively assessing their impact. However, existing techniques evaluate individual pixels rather than collective pixel groups forming patterns~\cite{park2022vision}, and lack uniform metrics for cross-image significance comparison.

\textbf{\circled{C3} Efficient Analysis and Summarization of Fake Patterns}: After identifying fake patterns constituted by influential pixel groups, the subsequent challenge is to devise an interface enabling users to effectively examine and summarize these patterns. The primary obstacle is the potential for an overwhelming number of patterns \cite{wang2023wizmap}, which could hinder clarity in analysis. At current, there are no established guidelines about how to design such an interface or tailor the user experience for such a workflow.

\textbf{\circled{C4} Adapting to a Diverse Range of Generative Models}: As GenAI technology rapidly evolves, it is critical that the analysis system adapts across various models. The system's design must also be flexible enough to accommodate new models as they emerge \cite{Ojha_2023_CVPR,cozzolino2023raising,zhu2023gendet}, ensuring its usefulness over time. This requires a modular approach, the specifics of which are yet to be determined.

\subsection{Design Goals}
\label{subsect:design_goals}

Next, to address \circled{C1}--\circled{C4}, we distilled a set of six design goals \circledVarD{G1}--\circledVarD{\rev{G5}}. The design and implementation of \sys, described in `\sys Backend' section and `\sys Interface' section, is intended to support these goals.

\textbf{\circledVarD{G1} Develop an interpretable classifier to detect real from fake images}: Discovering fake patterns (\circled{C1}) involves: (i) distilling critical information separating real from fake images into condensed representations, and (ii) leveraging these representations to detect key pixel groups. For this, we use supervised learning, by training a classifier not just to distinguish real from fake images, but also encode authenticity information~\cite{Ojha_2023_CVPR, cozzolino2023raising}. To handle diverse AI-generated images (\circled{C4}), we build the classifier on a large-scale pre-trained model with extensive training on varied datasets. 
\rev{We freeze the pre-trained  image encoder to leverage its generic visual feature extraction capabilities and train downstream classifiers on its image embeddings. To make the classifier interpretable, we introduce a distiller layer that maps these embeddings into a compact set of orthogonal authenticity-related dimensions, each of which can be linked to classifier weights and pixel-level relevance maps in subsequent design goals.}

\textbf{\circledVarD{G2} Using gradients to identify important pixels}: We leverage gradients propagated back from distilled representations to reveal output sensitivity to input changes~\cite{chefer2021generic}. \rev{Similar to gradient-based localization methods such as Grad-CAM~\cite{selvaraju2017grad}, this helps identify image regions that influence model behavior. However, rather than weighting convolutional feature maps using gradients from a final class logit, we adapt transformer relevance propagation for CLIP by propagating gradients from each dimension of the distilled authenticity representation to attention maps within CLIP's visual transformer blocks.} This yields dimension-specific token relevance maps that are mapped back to pixel sensitivities (\circled{C1}). Given the multi-dimensional distilled representation, we craft detailed masks for each dimension, as each uniquely encodes realness or artificiality information, creating comprehensive masks that trace specific pixel groups affecting authenticity.

\textbf{\circledVarD{G3} Establish a uniform metric for evaluating pattern influence}:
To quantify pattern influence (\circled{C2}), we address a limitation of gradient-based relevance maps, which primarily indicate relative importance within a single image~\cite{chefer2021generic}. \rev{Such maps reveal where influential pixels are located, but do not directly show how strongly a pixel group supports a ``real'' or ``fake'' prediction, nor whether that influence is comparable across images. We therefore aim to derive a signed contribution metric for each identified pixel group by linking it to the corresponding dimension in the distilled representation. The image-specific dimension value captures how strongly the evidence appears in the current image, while the classifier weight captures how strongly that dimension affects the real/fake decision globally. Combining and normalizing these factors yields a comparable score whose sign indicates prediction direction and whose magnitude indicates influence strength. This metric is intended to support visual comparison of pattern influence across images, groups, and prediction outcomes.}

\rev{\textbf{\circledVarD{G4} Support comparative analysis of prediction outcomes and recurring patterns}: Understanding fake patterns (\circled{C3}) requires comparing images and image groups rather than inspecting isolated predictions. Inspired by contrastive explanation, which helps explain why one outcome occurs rather than another~\cite{stepin2021survey}, users should be able to compare correctly classified and misclassified images that share apparent visual traits but differ in subtle authenticity-related patterns. Such comparisons should help users connect qualitative visual traits with quantitative representation differences, identify recurring patterns that distinguish successful detections from failures, and summarize how these patterns vary across real and fake image collections.}


\textbf{\circledVarD{\rev{G5}} Enable detailed examination of individual images}: Finally, in-depth analysis of fake images (\circled{C3}) likely requires examining individual examples of images of interest. This can be accomplished by providing an overview-plus-detail workflow, where users can navigate from a high-level summary view down to selecting and analyzing individual images at a fine-grained level.

%% file: documents/4-backend.tex

\section{\sys Backend}
\label{sect:backend}

We now describe \sys{}'s backend, which discovers patterns indicative of image authenticity or artificiality. This process involves training binary classifiers to distinguish between fake and real images, using these classifiers as encoders to transform images into interpretable vector representations, and then leveraging these representations to uncover relevant patterns. Specific implementation details are discussed the in following subsections.

\subsection{Datasets}
\label{subsect:real-and-ai-generated-datasets}

To train classifiers for \sys's pattern discovery, we utilize two state-of-the-art datasets comprising AI-generated images from leading generative model paradigms: GANs and diffusion models.

(i) The proGAN dataset~\cite{wang2020cnn} includes authentic images from the PASCAL object detection challenge and their fake counterparts generated by the proGAN model. This balanced dataset contains 30,000 real and 30,000 fake images per class across 20 classes. (ii) The DetectingSyntheticImage dataset\cite{corvi2023detection} focuses on images generated by latent diffusion models (LDMs) across various categories. We specifically examine 20,000 fake human face images generated by LDM, paired with real human face images from the Flickr-Faces-HQ (FFHQ) dataset, which contains 70,000 high-quality human face images.

For each generative model type (GAN and diffusion), we train a classifier to distinguish between ``real'' and ``fake'' images. The feature spaces of these trained classifiers are then utilized to identify deceptive patterns specific to each type of generated image.

To validate the classifier's performance and generalizability, we evaluate against state-of-the-art baselines~\cite{Ojha_2023_CVPR} using benchmark datasets, achieving comparable detection performance --- due to page constraints, please refer to the Appendix for benchmark performance details. Critically, while our classifier demonstrates competitive accuracy, we emphasize that our focus is also to support interpretability (which is not done by SOTA detectors). Put another way, our approach enables us to not only to identify AI-generated images, but also to provide interpretable insights into the specific patterns contributing to their detection, overcoming the black box problem intrinsic to many existing detectors.

\subsection{Learning Interpretable Representations}

\begin{figure*}[htbp!]
    \centering
    \includegraphics[width=.8\textwidth]{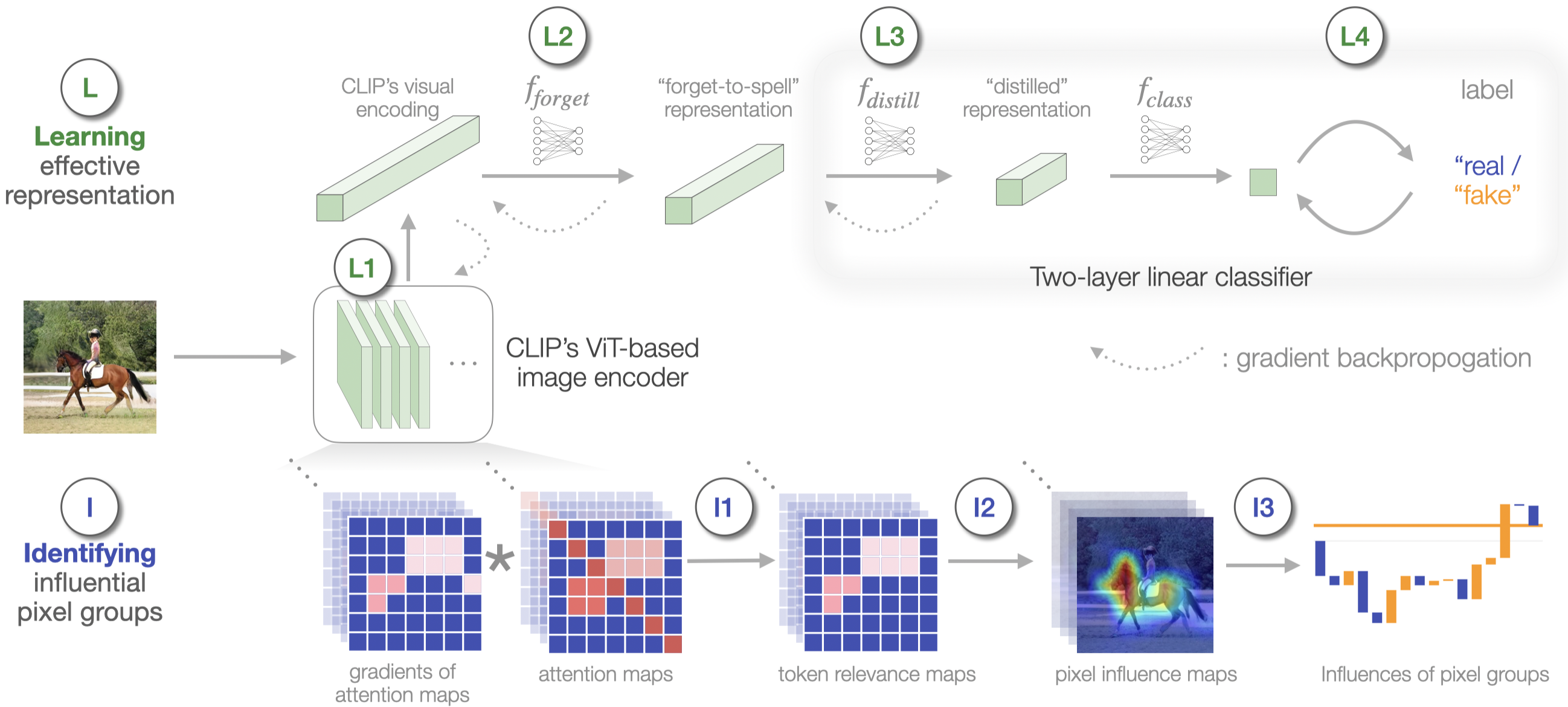}
    \caption{We employ a two-step approach for detecting AI-generated images and identifying their underlying patterns. Step L (Learning) begins by encoding images using CLIP's image encoder (L1) and then proceeds to strip textual information via a `forget-to-spell' projection (L2). Subsequently, we train a classifier that includes a distiller layer, which maps the `forget-to-spell' representation to a distilled space (L3), and a classification head that predicts the authenticity of images (real or fake) based on this distilled representation (L4). In Step I (Identification), we identify influential pixel groups by computing token relevance using a gradient-based technique (I1), translating token maps to individual pixel maps (I2), and computing uniform contribution metrics for pixel groups based on the classifier's weights (I3).}
    \label{fig:method}
\end{figure*}

At a high level, we are motivated by the fact that CLIP-based approaches have been shown effective for general-purpose fake detection, such as recent work by Ojha et al.~\cite{Ojha_2023_CVPR} Our approach is similar to the extent that we adapt CLIP for training a classifier. However, as discussed above, our intent is not only to achieve high classification performance but also to learn interpretable representations, which requires a customized approach. This process involves the following steps, and is summarized in \Cref{fig:method} L:

(i) \textbf{CLIP encoding}: The image is preprocessed using the CLIP:ViT-B/32 image encoder, $f_{\text{CLIP}}^{image}(\cdot)$, producing a 512-dimensional vector of generic visual features (L1 in \Cref{fig:method}).

(ii) \textbf{Text information removal}: A ``forget-to-spell'' projection, $f_{\text{forget}}(\cdot)$, filters out unintended text-related information from CLIP's embeddings. This is a key component in our approach (and also differentiates from prior techniques such as Ojha et al.~\cite{Ojha_2023_CVPR}) because, while CLIP's visual embeddings provide vital cues for differentiating between authentic and synthesized images~\cite{cozzolino2023raising}, 
its contrastive learning process inherently mixes visual and textual information~\cite{materzynska2022disentangling}. This mixing, though beneficial for tasks like zero-shot classification, can confound image authenticity verification. 
Our forget-to-spell projection uses an orthogonal projection technique~\cite{materzynska2022disentangling} and is trained with an orthogonality loss function on a diverse dataset (including natural images, text-only images, and images with overlaid text). This approach enables the model to distinguish visual elements from textual content, resulting in a more visually focused 256-dimensional vector $(f_{\text{forget}} \circ f_{\text{CLIP}}^{image})(I)$ (L2 in \Cref{fig:method}).

(iii) \textbf{Representation distillation}: A linear classifier trained on the forget-to-spell representation distinguishes between genuine and synthetic images. A distiller layer $f_{distill}(\cdot)$ reduces the 256-dimensional vector to 16 orthogonal dimensions (L3 in \Cref{fig:method}), followed by a classification head $f_{class}(\cdot)$ for final real/fake prediction (L4 in \Cref{fig:method}). (See `Representation Distillation' section below for details.)
   
The combination of $f_{CLIP}^{image}(\cdot)$, $f_{forget}(\cdot)$, and $f_{distill}(\cdot)$ forms an effective image encoder, converting images into a compact space rich with information crucial for determining image authenticity, thereby facilitating subsequent pattern discovery (\circledVarD{G1}).

\subsubsection{Representation Distillation}
\label{subsect:distill}
After obtaining visual-centric 256-dimensional representations $f_{forget}(f_{CLIP}^{image}(I))$, we further distill authenticity-indicative information through supervised learning. We train a binary classifier to distinguish real from fake images, guiding it to extract authenticity-relevant dimensions. This approach condenses 256 dimensions into 16 orthogonal dimensions (\rev{the authors empirically assessed this size as a good balance between performance and explainability by comparing against other potential values, though \sys can support different dimensional spaces and we plan to investigate tradeoffs in larger/smaller dimensional spaces as a future work}), enabling gradient-based analysis to identify influential pixel groups (i.e., patterns) for real/fake image classification.

Architecturally, our classifier is intentionally ``simple,'' only consisting of two linear layers: A distiller layer $f_{distill}(\cdot)$ reduces dimensionality retaining primarily image authenticity-relevant information, and a classification head $f_{class}(\cdot)$ processes the distilled representations through a sigmoid function to produce the final binary classification.
We train the classifier on the datasets described in `Datasets' section, initializing the distiller's weights orthogonally~\cite{saxe2013exact} and incorporating an orthogonality penalty in the loss function:
\vspace{-.5cm}

\begin{equation}
Loss(y,\hat{y}) = \lambda_{BCE} BCE(y,\hat{y}) + \lambda_{ortho} R(W)
\end{equation}

Here, $y$ is the image label, $\hat{y}$ is the classifier's output, and $R(W) = ||Id - WW^T||_2$ is the orthogonality penalty for the distiller's weight $W$ ($Id$ is the identity matrix). We set $\lambda_{BCE}$ to 3 and $\lambda_{ortho}$ to 1. The model is trained using Adam optimizer (batch size 32, learning rate 1e-3) for 12 epochs on a MacBook Pro with M1 chip, in 68 seconds.

This approach yields an encoder that condenses the forget-to-spell representations into a compact, 16-dimensional vector enriched with image authenticity information. We then apply a gradient-based method to these vectors to identify influential pixels and quantify their impact.

\subsection{Pixel Relevance Computation and Pattern Discovery}

Using the distilled representations, we identify influential fake image patterns by computing pixel relevances and selecting those with high relevance. Unlike conventional techniques that generate a single saliency map from final logits, we compute pixel relevance from each of the 16 dimensions in our distilled representation, producing 16 distinct masks, each highlighting unique influential regions: $\frac{\partial A}{\partial x_{i}}, i\in{1,...,16}$, where $A$ represents the attention map in CLIP's transformer block (\circledVarD{G2}). This approach has three key advantages:

(i) \textbf{Mask diversity}: Orthogonality among the 16 dimensions (see `Representation Distillation' section) ensures propagated gradients capture pixel importance from 16 distinct perspectives.

(ii) \textbf{Enhanced separability}: Our method isolates individual pixel regions and their contributions (unlike traditional aggregated saliency maps), supporting identification of influential areas.

(iii) \textbf{Fine-grained analysis}: The approach also enables detailed examination of \rev{individual} image regions (patterns), potentially surfacing subtle artificiality indicators often overlooked in black box analyses.

After identifying influential pixel groups, we quantify their impact using the classifier $f_{class}$'s global weights. This approach ensures consistent influence measurements across images, enabling direct cross-image comparisons. 
These relevance maps also facilitate segmentation and clustering of pixel groups into distinct visual concepts, enhancing analysis of common fake image patterns (\circledVarD{G4}). The discovered patterns are then presented in \sys{}'s interactive interface for user analysis and summarization (see `\sys Interface' section).

\subsubsection{Identifying Influential Pixel Groups}
\label{subsect:relevance_map}


To compute a pixel relevance map for identifying pixels crucial to the final real/fake prediction, we utilize gradients of attention maps in the transformer blocks within CLIP:ViT-B/32.

In ViT, an image $I$ is divided into $k \times k$ patches (where $k = 224 / 32 = 7$ for CLIP:ViT-B/32). These patches are tokenized into vectors, resulting in $k^2$ tokens per image, plus an additional [CLS] token, forming a $(k^2 + 1) \times d$ matrix, where $d = 512$ is the embedding space dimension. For an $h \times w$ image, we first compute a $k \times k$ token relevance map, $R_{k\times k}$, indicating each token's influence. This map is then upscaled to the original image size and normalized between 0 and~1. Initially, $R = [1]_{k\times k}$, considering all tokens equally contributory. We then refine R by backpropagating from the last transformer block:
\vspace{-.5cm}

\begin{equation}
R = R + \bar{A_{i}} R, i \in {m,...,1}
\end{equation}

Here, $\bar{A_i}$ is derived from gradients of attention maps in the $i$-th transformer block:
\vspace{-.5cm}

\begin{equation}
\bar{A_{i}} = \mathbb{E}_h\left( \nabla A_i \odot A_i \right)^+
\end{equation}

$\nabla A_i$ is the gradient of attention map $A_i$, and $\odot$ denotes the Hadamard product. Then, to focus on the most influential pixels, we simplify the final relevance computation to use only the last transformer block:
\vspace{-.5cm}

\begin{equation}
\label{eq:relevance}
R = \mathbb{E}_h\left( \nabla A_{last} \odot A_{last} \right)^+
\end{equation}

This method involves a forward pass to obtain the 16-dimensional distilled representation and attention maps, followed by dimension-wise gradient calculation. The resulting 16 relevance maps, each resized to the original image dimensions, serve as a set of masks highlighting different regions crucial for distinguishing real from fake images.

\subsubsection{A Uniform Metric for Pattern Contribution}
\label{subsect:pixel_contribution}

Based on the 16 relevance maps derived for each image, we introduce a method to quantify and compare contributions of different pixel groups. The original relevance scores in these maps are image-specific, preventing direct comparisons of pixel group influences across images. This limitation poses challenges in comparing pattern influences (\circledVarD{G3}, \circledVarD{G4}).
To address this, we develop a uniform metric that integrates both global and local dimensional information, yielding a normalized scalar between [-1, 1]. The scalar's sign indicates the pattern's contribution direction (positive for fake, negative for real), while its magnitude reflects the extent.

For an image $I$, we identify 16 pixel groups $pg_i, i \in {1,...,16}$, each corresponding to a dimension in the distilled representation $v_{distill} = (v_1, ..., v_{16})$. Using the classification head's weight $w = (w_1, ..., w_{16})$, we calculate $s_i = v_i \times w_i$ for each dimension. To enable cross-image comparisons, we normalize $s_i$ into a contribution score $c_i$:
\[c_i = \frac{|v_i \times w_i|}{\sum_{i=1}^{16} |v_i \times w_i|} * sign(\frac{v_i \times w_i}{\sum_{i=1}^{16} |v_i \times w_i|})\]

This preserves the sign of $s_i$ while avoiding potential amplification due to cancellations in summation. The resulting score encapsulates both local and global dimensional data for each pixel group, maintaining the original scalar's sign within the [-1, 1] range. This uniform metric also facilitates efficient comparison of pixel groups' influence on real/fake image classification both within and across images, providing a consistent measure of their impact.

\subsubsection{Revealing Commonality of Similar Images}
\label{subsect:visual_concept}

To illustrate commonalities among fake images exhibiting similar artificial patterns (\circledVarD{G4}), we leverage concept-based methods~\cite{kim2018interpretability}. Initially, we utilize the relevance maps generated in `Identifying Influential Pixel Groups' section as masks to identify influential image segments (pixel groups) within a collection of similar images. These images are considered similar based on the proximity of their ``distilled'' representations in the L2 distance metric. To refine our analysis, we exclude overlapping segments within images by discarding those with high Intersection over Union (IoU) scores relative to already identified segments, as they are nearly identical to the existing ones. Subsequently, we encode these segments using the combined encoder $(f_{distill}\circ f_{forget} \circ f_{CLIP}^{image})(\cdot)$ into distilled representations. We group these representations into three clusters (\rev{these were empirically assessed as a good balance in the number of groupings for our tested datasets, determined based on the number of images per cell}) for analysis using k-means clustering. 

This process enables us to identify and highlight clusters of visual concepts in a human-interpretable manner, effectively revealing the common visual patterns shared by groups of images with similar authentic or fake patterns.

\subsection{System Stack}
\rev{\sys{} is built on a Flask server to handle computations like interpretable image embeddings, influential pixel segmentation, and influence measurement in response to user queries. The frontend interface (see next section) is built with React and d3.js.}

%% file: documents/5-interface.tex

\section{\sys Interface}
\label{sect:interface}

\rev{Building upon our design goals (`Design Goals' section), techniques for extracting critical image authenticity information (`Representation Distillation' section), and methods for identifying influential pixel groups (`Identifying Influential Pixel Groups' section), we introduce \sys{}'s frontend interface. The system facilitates interactive and interpretable analysis and summarization of fake patterns in images produced by various generative models, across multiple levels of data granularity (i.e., global, subset, and local levels).}
At a high level, users begin using \sys{} by selecting a generative model and subject for comparison (e.g., real versus fake ``horses'' generated by the ``proGAN'' model) in a load menu. The main interface consists of four main views, shown in \Cref{fig:teaser} and described in detail below.

\subsection{Representation Overview} \label{subsec:representation_overview}
The Representation Overview (\Cref{fig:teaser}A) summarizes fake patterns, facilitates navigation and comparison, and supports user annotations. It groups similar image representations into distinct, non-overlapping cells to minimize visual clutter (\circledVarD{G4}), and uses a custom cell glyph (described below) that summarizes the detector's performance on the image subsets contained with each glyph, with the goal of intuitively presenting the proportions of undetected fake images. The overview plot is designed around the following design considerations:

\textbf{Avoiding visual clutters via local aggregation.} Using the combined encoder $f_{distill} \circ f_{forget} \circ f_{CLIP}^{image}$, we transform images into lower-dimensional distilled representations, placing images with similar patterns close together. To avoid overlap from traditional dimensionality reduction, we use IsoMatch~\cite{fried2015isomatch}, which preserves pairwise distances in a structured 2D grid layout. We divide the space into a user-definable $m \times m$ grid of non-overlapping cells (\textit{m} defaults to 30).
Deriving cells from distilled representations in this way ensures adjacent cells reflect similar real/fake information, supporting comparative analysis while also reducing clutter/overlap. Users can explore a cell and examine adjacent cells to recognize statistical differences and real/fake boundaries (e.g., where cells go from majority blue-to-orange).

\begin{figure}[t]
    \centering
    \includegraphics[width=.8\linewidth]{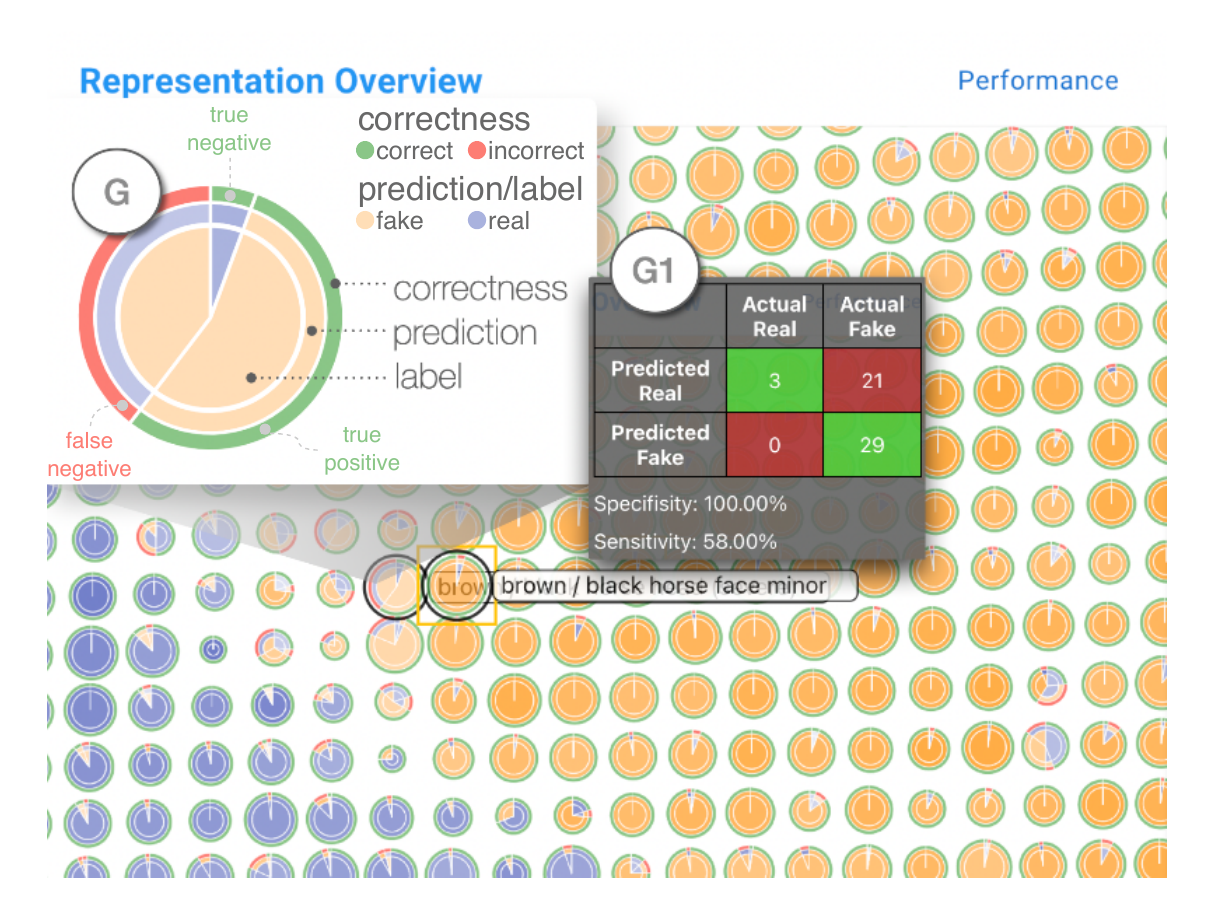}
    \caption{Our cell glyph design aggregates information for images within a cell. Users can investigate more detailed insights by hovering. Here, the tooltip shows details about a cell with severe 
    sensitivity (see \Cref{fig:violinplot}).
    }
    \label{fig:cell_glyph}
\end{figure}

\rev{\textbf{Supporting identification of suspicious cells with cell glyphs.}} \rev{We employ custom cell glyphs (see \Cref{fig:cell_glyph}) to help users quickly identify cells likely to contain deceptive or misclassified cases by visualizing the classifier's specificity and sensitivity. For this design, we initially considered matrix-like summaries of the four confusion outcomes, but because the overview contains many small cells, we prioritized rapid visual scanning over precise reading of individual values, which can be seen via a hover tooltip.} The glyph design is a pie chart with three radial layers and four sectors: \rev{true positives (fake images correctly predicted as fake), true negatives (real images correctly predicted as real), false positives (real images incorrectly predicted as fake), and false negatives (fake images incorrectly predicted as real)}. The outer arc uses consistent color coding: real images in blue, fake in orange, correct predictions in green, incorrect in red. Saturation indicates classifier confidence, with paler colors showing uncertainty; hovering displays a details-on-demand tooltip for exact values. \rev{The Representation Overview also supports navigation to better see glyph sectors and arcs, such as zooming and panning across glyphs along a blue-to-orange boundary in the IsoMatch grid.} Glyph size also reflects the number of examples, allowing users to quickly identify challenging examples near the decision boundary. Selecting a cell highlights it with a gold square and loads its contained images into linked panels (described below) for detailed analysis; users can pan and zoom to navigate.

\textbf{Annotation for incremental analysis.}  Users can right-click to annotate cells with pattern descriptions. Annotations persist until removed, enabling incremental understanding of fake patterns.

\subsection{Image View}
The Image View (\Cref{fig:teaser} B) activates when users select a cell, displaying its contained images and supporting tasks such as identifying shared fake patterns (\circledVarD{G4}). In this panel, displayed images are arranged in a 2D grid; we again utilize IsoMatch \cite{fried2015isomatch} for placing images in this grid (this means ``similar'' images will be placed close together). Each image is labeled (in its top right corner) with a circle: the color hue indicate the image's ground truth (blue for real, orange for fake); to the right and bottom of the image, a drop shadow shows its detector prediction. Discrepancies between circle label and drop shadow (e.g., for fake images wrongly identified as real) can indicate a need for more in-depth analysis.

Above, a ``Show Concept'' feature displays influential visual concepts in a Concept View popup (see \Cref{fig:concept_smile} for an example), using methods from `A Uniform Metric for Pattern Contribution' section, and intended to provide more insights into why misclassifactions might occur. For more detailed examination, users can select images via lassoing, which selects and loads them into the below-described Pattern View for fine-grained analysis of the critical pixels and potential fake patterns they contain.

\begin{figure*}[htbp!]
    \centering
    \includegraphics[width=.9\textwidth]{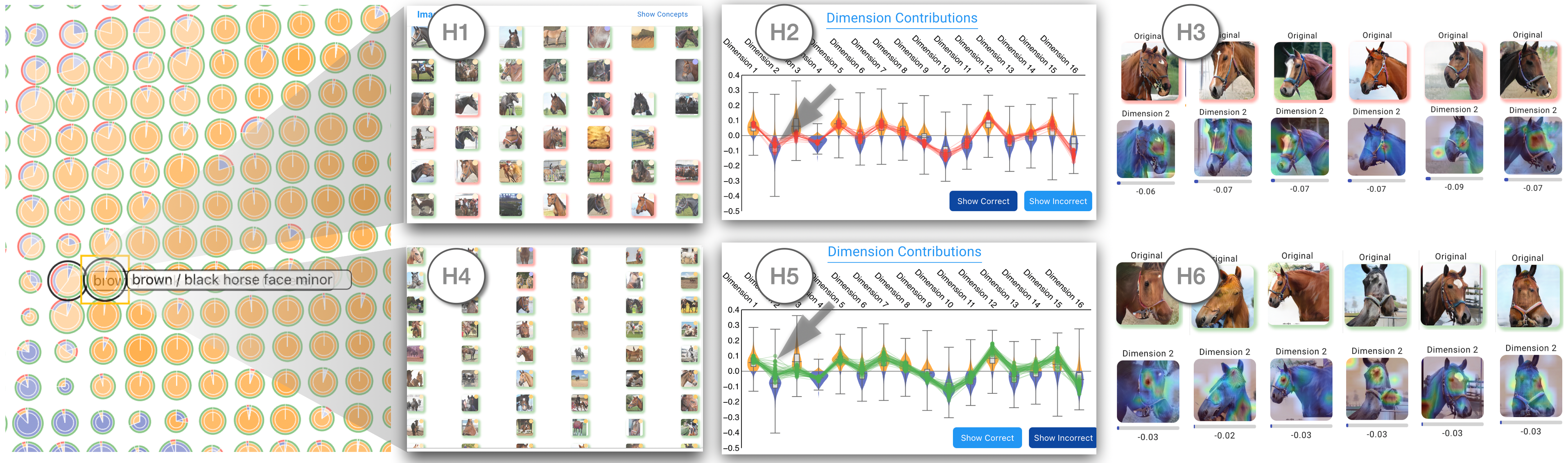}
    \caption{The user compares dimension contributions (H2 and H5) of two ``brown/black horse head'' cells with severe (H1) and minor sensitivity (H4). Analysis revealed notable differences in dimension 2. Investigation of pixel groups linked to dimension 2 showed strong correlation with pixels near horse eyes (H3 and H6). Severe sensitivity cell images had high dimension 2 values, while minor sensitivity images showed low values, suggesting horse eye portrayal significantly influences detection.}
    \label{fig:violinplot}
\end{figure*}

\subsection{Pattern View}

The Patten View (\Cref{fig:teaser} D) supports detailed examination of individual images (\circledVarD{\rev{G5}}), based on lassoing images in the Image View. The set of selected images is loaded into this view, which showcases the influential patterns of each selected image across the 16 distilled dimensions.

Specifically, each row shows information for a single image (when many images are loaded, users can scroll to see hidden rows), and each column information for one of the 16 dimensions. The original image is shown at the left-most position in the row (with drop shadow indicating prediction correctness), and then the image is duplicated 16 times across the row. Each of these images is overlaid with a heat map representing a pixel relevance map for its dimension (see `Identifying Influential Pixel Groups' section). Contribution bars below each heatmap show the dimension's influences on predictions (based on the methods in `A Uniform Metric for Pattern Contribution' section): orange for positive influences (pushing towards 'fake'), and blue for negative (pushing towards 'real'). For instance, in \Cref{fig:teaser} D, dimension 10 shows significant negative contribution for brown horse images, which misleads the classifier.

At each row's end, a waterfall chart summarizes the contribution of each dimension to the image's perceived authenticity. The chart exposes dimensions that are highly influential and the corresponding pixel groups, allowing users to identify the most significant patterns (E.g., in \Cref{fig:teaser}D, dimension 10 is key to predicting fake ``brown horse, side shot'' as real.) \erase{The waterfall chart also features a ``what-if'' button offering counterfactual analysis (\circledVarD{G4}), by presenting the minimal changes needed to alter prediction outcomes.}

\subsection{Dimension View}
To support \rev{contrastive} analysis from a quantitative perspective (\circledVarD{G4}), the Dimension View (\Cref{fig:teaser} C) allows users to examine quantitative attributes across cells, providing insights into each dimension's role. Users can toggle between \textit{dimension value plots} (shown in \Cref{fig:teaser} C) and a \textit{dimension contribution plot} (shown in \Cref{fig:violinplot} H2 and H5), revealing correlations between visual patterns and latent distributions.

Specifically, the dimension value plot uses small multiples to present dimension values across 16 dimensions. Upon initialization, it shows global distribution of real versus synthetic images. Dimensions are ordered by Kullback-Leibler divergence between real and fake distributions, with least divergence first, emphasizing dimensions with overlapping distributions requiring closer examination. For example, as shown in~\Cref{fig:teaser} C, dimension~1 shows nearly identical distributions. When selecting a cell, plots refresh to overlay the cell's distribution atop the global distribution, enabling users to gauge deviation from the global norm.

Toggling to the dimension contribution plot shows a set of violin plots with boxplot overlays showing key statistics on hover. This enables users to identify primary dimensions accounting for differences between visually similar images with divergent predictions (\circledVarD{G4}).
Upon selecting a cell, the plot incorporates parallel coordinates displaying each image's distilled representation. Users can use ``show correct'' and ``show incorrect'' buttons to filter contribution distributions, facilitating comparison to pinpoint critical dimensions contributing to prediction discrepancies (see \Cref{fig:violinplot}).

%% file: documents/6-usage-scenario.tex

\section{Usage Scenarios}
To help demonstrate the capabilities of \sys, we briefly present two usage scenarios from the perspective of a hypothetical user \user, who has to analyze various synthetic patterns produced by different generative models. The first scenario focuses on horse images created by the proGAN model, while the second investigates synthetic human faces generated by the LDM.
\rev{In these scenarios, the analyst works with a curated dataset in which images are already grouped by generator and semantic class, and therefore begins by focusing on a specific subject (e.g., horses or human faces).}

\textbf{Scenario 1: Analyzing Patterns in proGAN Images.}
\user begins by selecting the proGAN model and ``horse'' as the subject in the interface. Upon loading, the Representation Overview (\Cref{fig:teaser} A) displays images organized into local cells. \user notices a pale central region suggestive of a decision boundary, surrounded by numerous orange cells with red arcs, indicating deceptive synthetic patterns (\circledVarD{G4}).
He selects a prominent orange cell in the center, revealing multiple synthetic images of brown horses in Image View (\Cref{fig:teaser} B). Many of these images are misclassified as real. \user tags this cell as ``brown horse, side shot'' and explores adjacent cells with similar content (\circledVarD{G4}). Further exploration leads \user to a cell with a high misclassification rate featuring horse heads, which he tags as ``brown/black horse head'' (this cell's glyph is shown in \Cref{fig:cell_glyph}).

For a more detailed analysis, \user investigates the ``brown/black horse head'' pattern by comparing two adjacent cells containing similar images but with notably different accuracy rates (\Cref{fig:violinplot} H1, H4). Using the``Show Concepts'' feature in the Concept View, he identifies shared key visual concepts within the images, such as fur and sky patterns. He then refers to the Dimension Value Plot and observes that the dimension distributions for these two cells are quite similar across most dimensions, indicating that their differences are subtle. To reveal these subtle differences, \user examines the Dimension Contribution Plot (\Cref{fig:violinplot} H2, H5), which reveals a significant difference in dimension~2's contribution between the two cells. The Pattern View (\Cref{fig:violinplot} H3, H6) further highlights that the pixel groups linked to dimension~2, which correlate with areas near the horse's eyes, significantly impact the classification (\circledVarD{G5}). Put another way, \user has discovered that horse eyes and ears --- even when appearing similar between real and fake images --- can nonetheless play a key role in (mis)detection of fake images based on their latter pixel patterns. More broadly, these detailed analyses demonstrate how \sys enables users to uncover subtle-yet-significant factors that contribute to the classification of synthetic images, via supporting both pattern identification and \rev{interpretation with quantitative model-derived cues}.

As an addendum to this scenario, when zooming out, \user discovers a red-bordered blue cell contained amidst orange cells, which indicates real images were confusingly misclassified as fake. We report his analysis on this cell (which predominantly contains real horse images in 3D video games) in the Appendix.

\begin{figure}[htbp!]
    \centering
    \includegraphics[width=\linewidth]{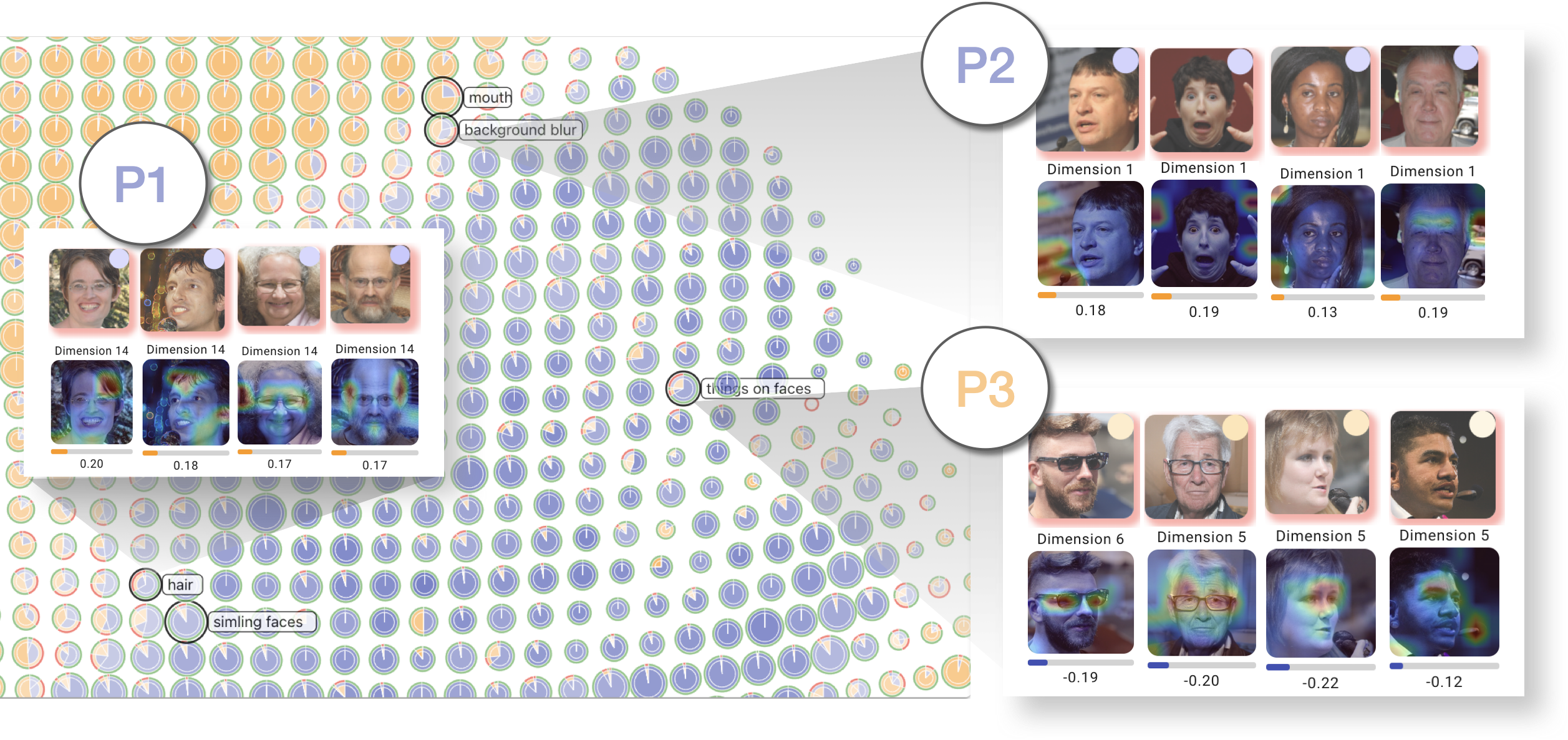}
    \caption{The user identifies common patterns in misclassified real faces, including hair (P1) and background blur (P2), as well as factors contributing to the misclassification of fake images, such as ``things on faces'' (P3).}
    \label{fig:hf_patterns}
\end{figure}

\begin{figure}[t]
    \centering
    \includegraphics[width=\linewidth]{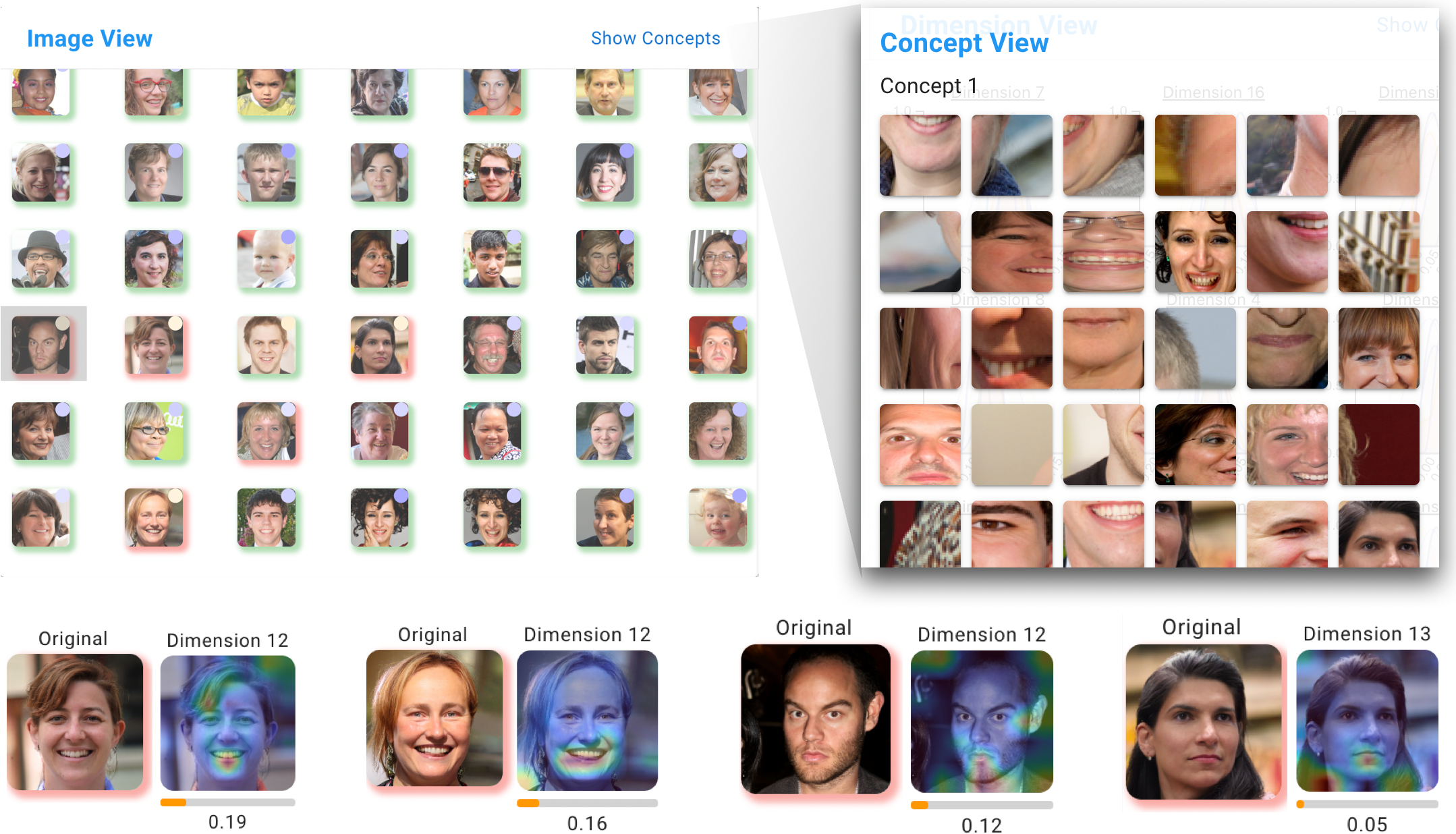}
     \caption{The user identifies a cell with misclassified images, mostly smiling faces. The Concept View reveals that the mouth significantly influences classification. Despite this, the smiles are correctly detected by the model as they have a positive contribution, indicating the classifier recognizes ``fake smiles.'' \erase{Further analysis shows the misclassification is due to high positive contributions from the forehead region.}}
    \label{fig:concept_smile}
\end{figure}

\textbf{Scenario 2: Analyzing Patterns in LDM Images.}
\user now switches to the LDM, with ``human face'' as the subject. The representation overview (shown in \Cref{fig:hf_patterns}) again shows a clear decision boundary between real and synthetic images.
He tags two noteworthy cells: one with low sensitivity (true positive rate) containing misclassified synthetic images featuring ``things on faces,'' and the another with images of smiling individuals, which he labels ``smiling faces.'' Based on this, he conducts a more detailed analysis of these two patterns:

For the ``things on faces'' pattern, images in this cell (\Cref{fig:hf_patterns} P3) show human faces with realistic items like sunglasses, eyeglasses, and microphones. These near-face objects significantly influence the classifier to predict these images as real (\circledVarD{G4}, \circledVarD{G5}).
For the ``smiling faces'' pattern, the Concept View (\Cref{fig:concept_smile}) highlights the mouth region's influence on classification.
Surprisingly, the mouth region in misclassified images contributes to a ``fake'' classification.
Further investigation reveals that misclassifications often stem from the forehead region (\circledVarD{G4}).

In addition to analyzing fake image patterns, \user also investigates real images that are flagged as synthetic (\circledVarD{G4}). By examining cells with high false-negative rates (predominantly blue with red borders), he identifies recurring misclassification patterns: real images are often misjudged based on hair features associated with dimension 14 (\Cref{fig:hf_patterns} P1) or background blur associated with dimension 1 (\Cref{fig:hf_patterns} P2) (\circledVarD{G4}). These insights points to the classifier's vulnerabilities, suggesting that to compromise an image's perceived authenticity, one might only need to introduce specific patterns to trigger a false positive classification.

%% file: documents/7-user-study.tex

\section{User Study}
\label{sec:user_study}
To empirically evaluate \sys{}, we conducted a user study with 10 participants. The study aimed to answer two primary questions: (i) How well does the system support summarizing patterns of fake images (\circledVarD{G1}--\circledVarD{G2})? (ii) How well does the system support \rev{analysis of specific fake image patterns using quantitative model-derived cues} (\circledVarD{G3}--\circledVarD{G5})?
Both quantitative and qualitative data were collected to assess the system's ability to support the design goals and its overall usability. \rev{This study was performed under the oversight of the author's institutional review board, and we obtained informed consent from all participants prior to beginning the study.}

\subsection{Study Design}
Participants followed a three-stage procedure:

(1) \textbf{Training:} After completing a demographics questionnaire, participants watched a 5-minute video explaining \sys's functionality and interface, with opportunities to ask questions. Participants then completed a training task identifying a fake pattern in bird images and reasoning about contributing dimensions to familiarize themselves with the system.

(2) \textbf{Tasks:} Participants completed two tasks designed to evaluate key aspects of the system:

\textbf{T1: Identify fake patterns.} This task assessed summarization capabilities (\circledVarD{G1}, \circledVarD{G2}, \circledVarD{G4}). Participants identified five fake patterns in two image classes: horses (proGAN-generated, easier) and human faces (LDM-generated, harder), without time constraints, though completion time was recorded to measure efficiency. A think-aloud protocol captured differences in mental workload between the datasets, \rev{and task correctness was assessed based on accurately identifying and describing actual patterns to the study proctor.}

\textbf{T2: Analyze a specific fake pattern.} This task evaluated \rev{participants' ability to use quantitative model-derived cues to interpret a specific fake pattern} (\circledVarD{G3}, \circledVarD{G4}). 
Participants analyzed the ``person-on-horse'' pattern and specified  relevant dimensions, with completion time recorded. \rev{Task correctness was assessed based on the accurately interpreting the pattern and describing its relevant dimensions to the study proctor.}

(3) \textbf{Review:} \rev{After completing both tasks,} participants completed a usability survey, rating system components using 7-point Likert scales and providing comments.

\subsection{Participants and Apparatus}
\rev{We recruited ten participants: eight CS graduate students and one post-doc, and one graduate student in education technology (mean age = 28.4, SD = 2.6; 7 males, 3 females). All participants had experience with generative image AI for image creation using natural language prompts (e.g., via Dall-E or MidJourney) and all were computationally proficient at image review and editing (e.g., via Photoshop), and self-rated as competent at using visual analytics for data-driven analysis. 
Study sessions were conducted via Zoom (avg. 35 min, SD = 12).
}

\subsection{Study Results}
For T1, average completion time was 8 minutes (SD = 1). For T2, average completion time was 18 minutes (SD = 10), though this was expected due to its open-ended and reflective nature (e.g., exploring and reasoning about patterns). All participants successfully completed both tasks by providing the requested pattern labels and dimension-based explanations. We report quantitative ratings from the usability survey and qualitative verbal comments collected via the think-aloud protocol and summary answers. Verbal comments were thematically coded using an open coding approach \cite{thornberg2014grounded}, focusing on how the system promotes insights and supports \circledVarD{G1}--\circledVarD{G5}.

\begin{figure}[t]
\centering
\includegraphics[width=\linewidth]{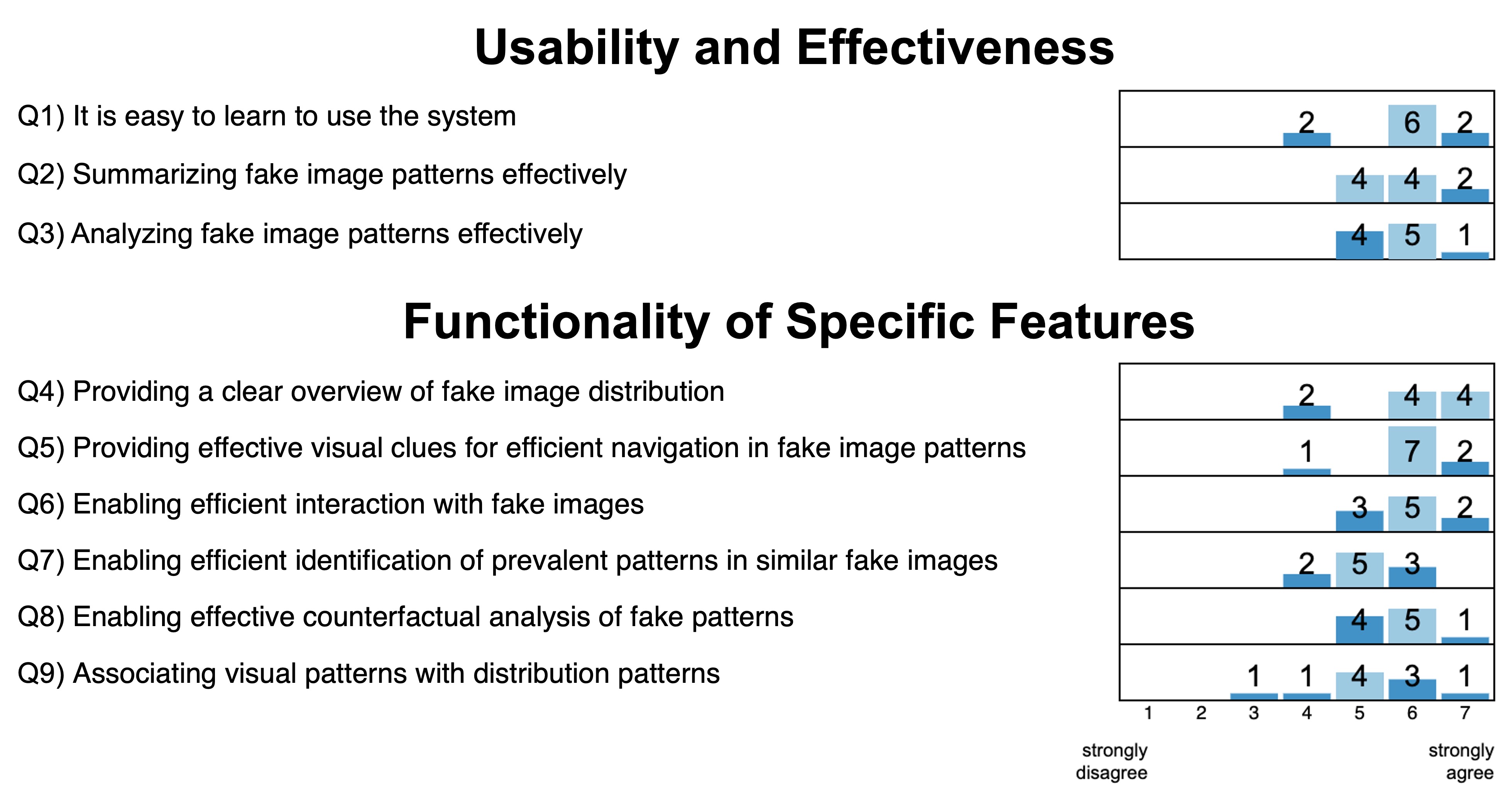}
\caption{Participant usability ratings of \sys from the survey during the user study review stage. Median ratings are shown in brighter blue.}
\label{fig:study_ratings}
\vspace{-.65cm}
\end{figure}

\textbf{System Usability and Functionality Ratings.}
\Cref{fig:study_ratings} summarizes usability and functionality ratings from the review stage survey in two categories: (Q1--Q3) usability and effectiveness, and (Q4--\rev{Q9}) functionality of specific features. Feedback was overall positive, with the system rated as easy to use (Q1), effective at summarizing fake patterns (Q2), and supporting analysis (Q3). The representation overview panel (Q4--Q5), image view (Q6--Q7), pattern view (Q8), and dimension view (Q9) likewise received high ratings.

\textbf{User Feedback.}
Participant verbal comments provided additional insights into the types of insights users can generate, in the context of design goals. We summarize several themes below:

\textbf{(\circledVarD{G1}, \circledVarD{G2}) Representation Overview enables efficient navigation and exploration.} The Representation Overview and Image View were positively regarded for navigating the image space and exploring fake patterns. Participants tended to first inspect large cells with major orange portions bordered by red, indicating high numbers of misclassified fakes, then use the Image View for more fine-grained review. ``\textit{I'm starting on cells with a lot of misclassified fake images}'' (p4). ``\textit{This [overview] makes it easy to browse the data}'' (p2). ``\textit{I like the efficiency [of the representation overview]. I can check a lot of images at one time}'' (p9).

\textbf{(\circledVarD{G4}, \circledVarD{\rev{G5}}) Pattern View allows in-depth analysis of specific fake patterns.} Participants found the Pattern View effective for analyzing fake patterns by highlighting influential pixel groups, measuring their impact, and enabling focused analysis. ``\textit{I like this view. You can see the detector checks the horse and doesn't have an issue there, but it also checks other pixels in the background and finds the fakeness. So it knows the fake pattern among these images comes from the background}'' (p4). ``\textit{It's clear the neural net picks up the person [pattern]. You see it looks at the person [rider on horse]. When the person is realistic, it passes [i.e., predicts true], and when the person is fake, it predicts fake}'' (p8).

\textbf{(\circledVarD{G3}, \circledVarD{G4}) Dimension View allows quantitative association of fake patterns with dimension distributions.} Participants were able to link fake patterns with dimension distributions. ``\textit{I can see that d8 is for the person [pattern]. It's positive [indicating fake] when it sees a fake person and near zero when it doesn't}'' (p6, analyzing the ``rider-on-horse'' pattern in the horse class). ``\textit{d3 is a background dimension, it looks at the background and it has very little bias}'' (p7, analyzing the ``rider-on-horse'' pattern in the horse class). ``\textit{You can see d16 and d10 are for the horse body}'' (p10, analyzing the ``brown horse'' pattern).
Some potential improvements were suggested here, such as enlarging images and providing natural language descriptions of patterns. We \revsec{intend} to explore these refinements as future work.

\rev{These comments also aligned with our observations of system use during the study. For example, participants tended to consistently identify the more salient patterns during their work (e.g., the above-discussed rider-on-horse pattern), though finer-grained patterns varied across individuals.}

%% file: documents/8-discussion_and_future_work.tex

\section{Discussion and Conclusion}

ASAP \rev{addresses three major challenges currently faced in fake image analysis workflows:}
limited \textit{generalizability}, lack of \textit{interpretability}, and poor \textit{\rev{actionability}}.
For generalizability, we develop a detector that leverages pre-trained models and supports seamless integration of future models. For interpretability, we use an interpretability-focused training paradigm that distills interpretable embeddings and introduces novel techniques for segmenting influential pixel regions and assessing influence. This uncovers latent, recurrent patterns in AI-generated images. An \rev{actionable} interface organizes these patterns, \rev{enabling users to investigate and reason about the causes of detector successes and failures.}

Overall, our usage scenarios and user study demonstrate \sys's effectiveness in supporting design goals (\circledVarD{G1}--\circledVarD{G5}). Based on our process of  developing and validating \sys, we also identified several areas for future visual analysis 
and explainability of GenAI images and models:

\textbf{Limits in detector capacity:}
Our detector's capacity is currently bound by the CLIP architecture. However, our pipeline's flexible design allows for integration of alternative or emerging detectors, \rev{imposing these two requirements: (i) there should be a differentiable feature space from which the lightweight forget-to-spell projection and distiller layer can be trained, and (ii) relevance maps should be computable. For a transformer-based encoder, swapping in a new backbone would mainly require retraining the two small linear layers on top. For non-transformers, requirement (ii) would need to be replaced with an architecture-appropriate attribution method (e.g., a CNN-style saliency technique), which is a more involved adaptation.}

\textbf{Real-time labeling and evaluation:} Our current labeling method operates in a post-hoc manner, focusing on interactively and interpretably support AI-generated content analysis where ground truth image labels are known. While this makes \sys highly effective as a tool for analyzing image datasets that contain mixtures of AI-generated and real images, future iterations can adapt the system for scenarios where users want to analyze an image's authenticity in real-time (i.e., without knowing its true label).

\rev{\textbf{Scalability considerations:}
While \sys{} supports analyzing analysis at multiple granularities (from the overall image dataset in the Representation Overview down to individual image details in the Pattern and Dimension Views), we note the current system design imposes some limitations on scalibility. For one, Isomatch has poor computational complexity (O($n^3$)), which we address by pre-computing for the default layout for the Representation View grid to maintain real-time interaction. However, for users who wish to explore changing the grid size, especially for larger image ensemble datasets, this could lead to a computation bottleneck. In such cases, alternative layout algorithms could be explored, though it would be important that the results still supported workflows such as being able to recognize real/fake boundaries where cells go from majority blue-to-orange.}

\rev{Another challenge for the Representation View is that, when a glyph is selected, all of its contained images are shown in the Image View. In our tested datasets, each glyph usually aggregated to at most 75 images, though this size would increase if the Representation View's grid size was shrunk or an image dataset containing millions of images was tested. In such cases, the Image View would likely need to be redesigned to better support larger amounts of selected images.
}

\rev{\textbf{Limits of saliency-based explanations:} While our relevance maps help identify image regions that influence predictions, they do not by themselves explain why those regions are influential or what specific visual attributes within them drive the model's decision. For example, a highlighted mouth region may indicate that the area is important for predicting an image as fake, but not whether this is due to shape, texture, boundary artifacts, or some other property. Richer explanations of the specific attributes driving model behavior remain an important direction for future work.}

\textbf{Interactive base model selection:} While \sys currently employs CLIP as its foundational visual feature encoder, the rapid advancement of multi-modal pre-trained models presents an opportunity to support interactive base model selection. This enhancement could significantly improve the detection of nuanced patterns in AI-generated images.

\textbf{Integration with LLMs:} As multi-modal LLMs become more powerful, there is potential to integrate into them deepfake analysis workflows, such as using them to verbally summarize fake patterns. This could provide users with a more efficient overview of fake patterns and their distribution. However, given the current limitations in LLMs' vision understanding capabilities~\cite{zhang2024exploring}, this integration will likely require more mature models (or advanced RAG approaches) than are currently available.

\rev{\textbf{Ethical and privacy considerations:} Finally, deepfake analysis tools such as \sys{} can also raise ethical concerns, due to their potential dual-use nature: while they can support media forensics and authenticity analysis, discovered patterns could also be misused to help deceptive content evade detection. In addition, scenarios involving human face images introduce privacy concerns when facial content is inspected and compared at scale, even if (as in our case) public benchmark datasets are used. More broadly, generative ecosystems continue to face concerns related to bias and copyright, motivating not only calls for responsible use but also the development of guardrails to constrain unintended outcomes.}



Ultimately, \sys presents a novel approach for detecting patterns and signatures in AI-generated images and facilitating their in-depth analysis. 
In particular, \sys addresses key challenges in existing approaches by supporting generalizability, interpretability, and \rev{actionability}, via a combination of backend techniques and interactive, human-in-the-loop visual analytics. While \sys currently focuses on image content, as the landscape of GenAI continues to rapidly evolve, we posit systems like \sys will play an increasingly crucial role in supporting digital media integrity and fostering trust in content.

%% file: documents/appendix.tex

\section{Validating the Generalizability of the Detector}

We evaluate our detector's generalizability following the methodology of Ojha et al.~\citeappx{Ojha_2023_CVPR}. Our approach involves training the detector on the proGAN dataset, which contains real images from ImageNet and fake images generated by the proGAN model~\citeappx{wang2020cnn}. We then test the detector on unseen datasets to assess its generalizability.

In \Cref{tab:accuracy_comparison,tab:precision_comparison}, we report the performance of various detection methods across different image generation techniques. The models included are:

\begin{itemize}
    \item GAN-based: CycleGAN~\citeappx{zhu2017unpaired}, BigGAN~\citeappx{brock2018large}, StyleGAN~\citeappx{karras2019style}, GauGAN~\citeappx{park2019semantic}, and StarGAN~\citeappx{choi2018stargan}
    \item Diffusion-based and hybrid techniques~\citeappx{rombach2022high, dhariwal2021diffusion} 
    \item Other generative technologies~\citeappx{rossler2019faceforensics++, chen2018learning, dai2019second, li2019diverse}
\end{itemize}

We report both accuracy and precision for each method. Accuracy represents the model's ability to make correct predictions overall, while precision indicates its ability to correctly identify fake images.

Our method outperforms many of the baseline models, and is comparable to that of Ojha et al.~\citeappx{Ojha_2023_CVPR}, another state-of-the-art model that also leverages CLIP. Differences in performance are likely attributed to our forget-to-spell projection and distillation process, which is needed for interpretability purposes (and which Ojha et al. lacks). In our case, this process necessarily discards some information from the CLIP embeddings, resulting in slightly less informative input for detection. However, as discussed in the main paper body, our primary goal is to discover fake patterns rather than maximize prediction accuracy. This trade-off is thus acceptable given the overall comparable performance.

\section{Patterns of Confusing Real Horse Images}
\label{appendix:real_horse_images}
At the end of Usage Scenario 1, hypothetical user Ryan finds a cell containing real horse images that are confusingly misclassified as synthetic/fake. 
These are predominantly real horse images from 3D games, and dimension~6 (associated with the horse's body) contributes significantly to their synthetic classification, which suggests that 3D-generated horses lack certain realistic features.

\begin{figure}[t]
\centering
\includegraphics[width=\columnwidth]{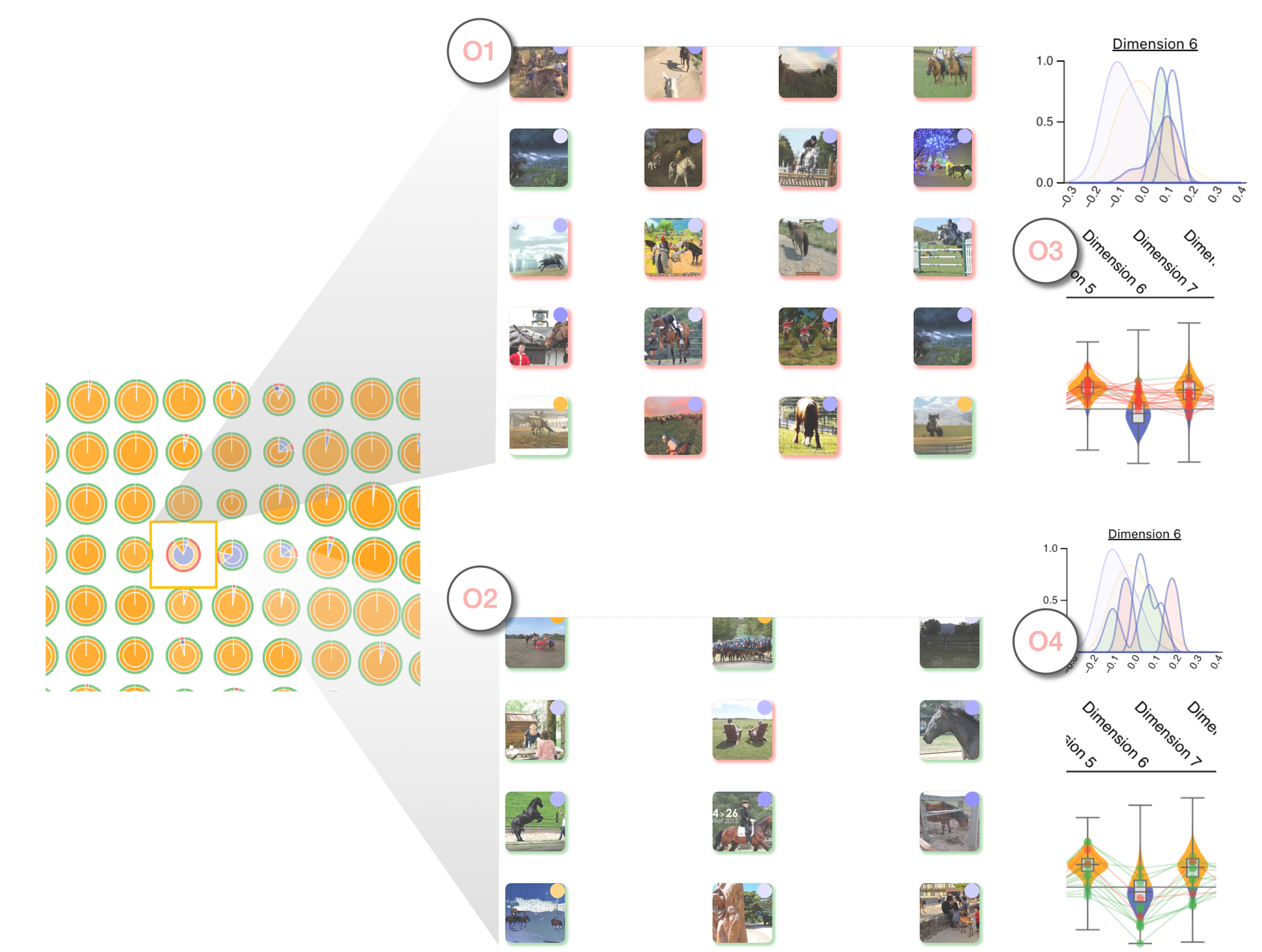}
\caption{Real horse images from 3D games misclassified as synthetic (red-bordered cell) and nearby cells.}
\label{fig:real-horses1}
\end{figure}

To reach such a finding, \Cref{fig:real-horses1} shows Ryan interacting with the cell glyph in the Representation Overview.
Here, Ryan first observes that most images in one cell (O1) depict horses in 3D games. A nearby cell (O2) also contains mostly real images, but these are correctly classified. Both cells' images feature horses with grassy backgrounds.

\begin{figure}[t]
\centering
\includegraphics[width=\columnwidth]{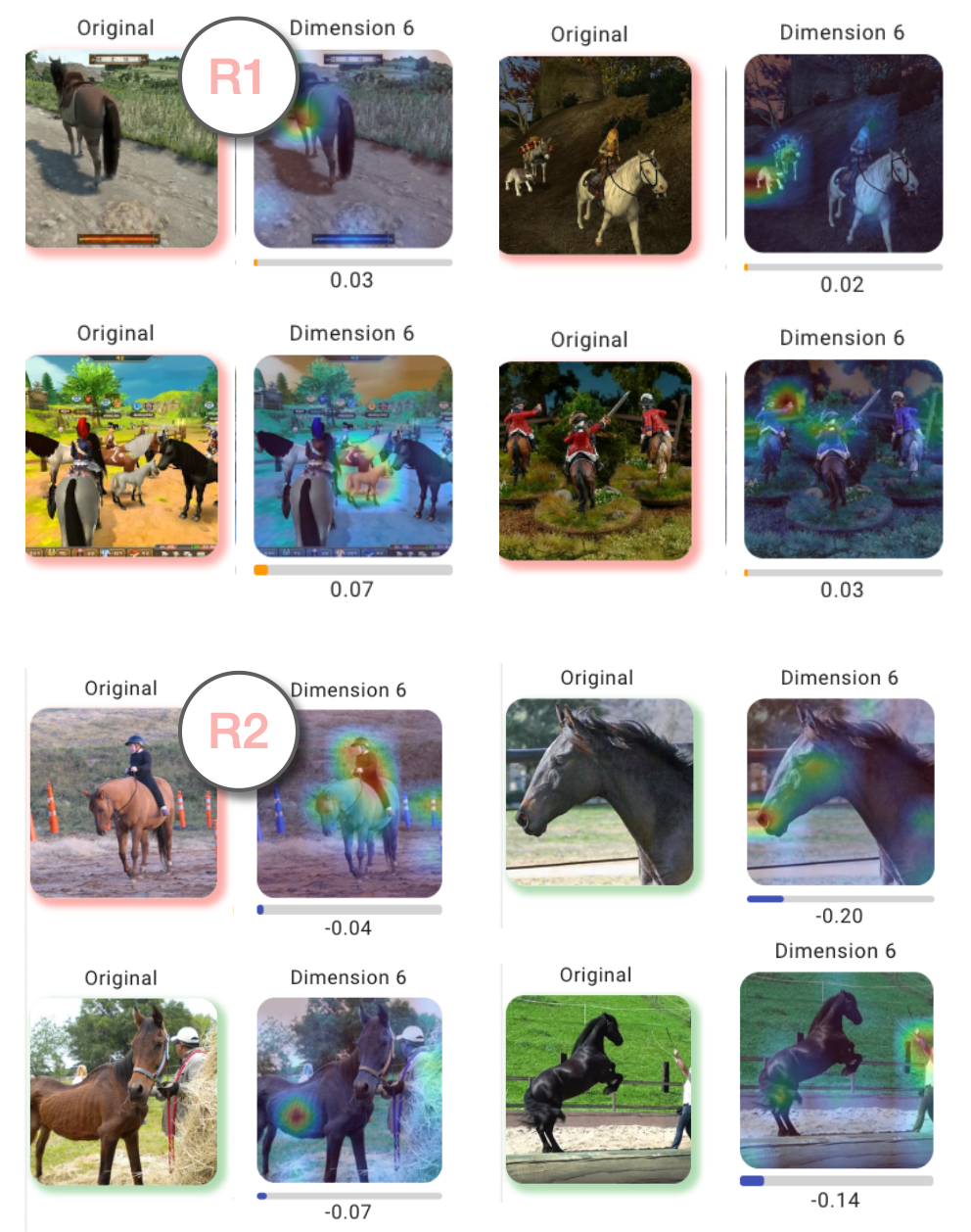}
\caption{Dimension 6 focuses on pixels near the horse body or the rider. It correctly predicts real images (assign negative contributions) in cell O2 but fails for cell O1 (assign positive contributions).}
\label{fig:real-horses2}
\end{figure}

By comparing these cells, their dimension values, and dimensional contributions, Ryan identifies dimension 6 as the key differentiator (O3, O4).
\Cref{fig:real-horses2} illustrates that dimension 6 focuses on pixels near the horse's body or the rider. While this dimension successfully predicts real images in cell O2, it fails to do so for cell O1.

\onecolumn
\begin{table*}[!hbp]
  \centering
  \resizebox{\textwidth}{!}{
    \begin{tabular}{c*{20}{c}}
    \toprule
    \multirow{2}{*}{\shortstack[c]{Detection\\Method}} & \multicolumn{7}{c}{Generative Adversarial Networks} & \multicolumn{2}{c}{Low-level Vision} & \multicolumn{2}{c}{Perceptual Loss} & \multirow{2}{*}{Guided} & \multicolumn{3}{c}{LDM} & \multicolumn{3}{c}{Glide} & \multirow{2}{*}{DALL-E} & \multirow{2}{*}{\shortstack[c]{Avg.\\Acc.}} \\
    \cmidrule(lr){2-8} \cmidrule(lr){9-10} \cmidrule(lr){11-12} \cmidrule(lr){14-16} \cmidrule(lr){17-19}
    & ProGAN & CycleGAN & BigGAN & StyleGAN & GauGAN & StarGAN & Deepfakes & SITD & SAN & CRN & IMLE & & 200 steps & 200 w/ CFG & 100 steps & 100/27 & 50/27 & 100/10 & & \\
    \midrule
    Wang et al. \citeappx{wang2020cnn} & 99.99 & 85.20 & 70.20 & 85.7 & 78.95 & 91.7 & 53.47 & 66.67 & 48.69 & 86.31 & 86.26 & 60.07 & 54.03 & 54.96 & 54.14 & 60.78 & 63.8 & 65.66 & 55.58 & 69.58 \\
    Chai et al. \citeappx{chai2020makes} & 75.03 & 68.97 & 68.47 & 79.16 & 64.23 & 63.94 & 75.54 & 75.14 & 75.28 & 72.33 & 55.3 & 67.41 & 76.5 & 76.1 & 75.77 & 74.81 & 73.28 & 68.52 & 67.91 & 71.24 \\
    Nataraj et al. \citeappx{nataraj2019detecting} & 97.70 & 63.15 & 53.75 & 92.50 & 51.1 & 54.7 & 57.1 & 63.06 & 55.85 & 65.65 & 65.80 & 60.50 & 70.7 & 70.55 & 71.00 & 70.25 & 69.60 & 69.90 & 67.55 & 66.86 \\
    Zhang et al. \citeappx{zhang2019detecting} & 49.90 & 99.90 & 50.50 & 49.90 & 50.30 & 99.70 & 50.10 & 50.00 & 48.00 & 50.60 & 50.10 & 50.90 & 50.40 & 50.40 & 50.30 & 51.70 & 51.40 & 50.40 & 50.00 & 55.45 \\
    Ojha et al. \citeappx{Ojha_2023_CVPR} & 100.0 & 98.50 & 94.50 & 82.00 & 99.50 & 97.00 & 66.60 & 63.00 & 57.50 & 59.5 & 72.00 & 70.03 & 94.19 & 73.76 & 94.36 & 79.07 & 79.85 & 78.14 & 86.78 & 81.38 \\
    \rowcolor[gray]{0.9} \textbf{Ours} & 97.45 & 91.14 & 71.33 & 82.15 & 78.01 & 91.92 & 67.12 & 64.17 & 69.18 & 65.02 & 73.28 & 79.95 & 79.00 & 63.50 & 80.50 & 87.00 & 88.55 & 88.10 & 60.60 & 77.79 \\
    \bottomrule
    \end{tabular}
  }
  \caption{\textbf{Generalization Results (Accuracy).} This table compares the classification accuracy of various methods for detecting real and fake images. Our approach achieves performance ahead of most existing methods and is comparable to the state-of-the-art method \protect\citeappx{Ojha_2023_CVPR} with a slight difference attributable to our focus on interpretability. Our detector incorporates an orthogonality loss during training, which regularizes the model's weights. This design choice balances classification performance with the generation of orthogonal dimensions in the distiller layer, facilitating subsequent relevance-based mask generation and fake pattern discovery.}
  
  \label{tab:accuracy_comparison}
\end{table*}

\begin{table*}[!hbp]
  \centering
  \resizebox{\textwidth}{!}{
    \begin{tabular}{c*{20}{c}}
    \toprule
    \multirow{2}{*}{\shortstack[c]{Detection\\Method}} & \multicolumn{7}{c}{Generative Adversarial Networks} & \multicolumn{2}{c}{Low-level Vision} & \multicolumn{2}{c}{Perceptual Loss} & \multirow{2}{*}{Guided} & \multicolumn{3}{c}{LDM} & \multicolumn{3}{c}{Glide} & \multirow{2}{*}{DALL-E} & \multirow{2}{*}{\shortstack[c]{Avg.\\mAP}} \\
    \cmidrule(lr){2-8} \cmidrule(lr){9-10} \cmidrule(lr){11-12} \cmidrule(lr){14-16} \cmidrule(lr){17-19}
    & ProGAN & CycleGAN & BigGAN & StyleGAN & GauGAN & StarGAN & Deepfakes & SITD & SAN & CRN & IMLE & & 200 steps & 200 w/ CFG & 100 steps & 100/27 & 50/27 & 100/10 & & \\
    \midrule
    Wang et al. \citeappx{wang2020cnn} & 100.0 & 93.47 & 84.5 & 99.54 & 89.49 & 98.15 & 89.02 & 73.75 & 59.47 & 98.24 & 98.4 & 73.72 & 70.62 & 71.0 & 70.54 & 80.65 & 84.91 & 82.07 & 70.59 & 83.58 \\
    Chai et al. \citeappx{chai2020makes} & 80.88 & 72.84 & 71.66 & 85.75 & 65.99 & 69.25 & 76.55 & 76.19 & 76.34 & 74.52 & 68.52 & 75.03 & 87.1 & 86.72 & 86.4 & 85.37 & 83.73 & 78.38 & 75.67 & 77.73 \\
    Nataraj et al. \citeappx{nataraj2019detecting} & 99.74 & 80.95 & 50.61 & 98.63 & 53.11 & 67.99 & 59.14 & 68.98 & 60.42 & 73.06 & 87.21 & 70.20 & 91.21 & 89.02 & 92.39 & 89.32 & 88.35 & 82.79 & 80.96 & 78.11 \\
    Zhang et al. \citeappx{zhang2019detecting} & 55.39 & 100.0 & 75.08 & 55.11 & 66.08 & 100.0 & 45.18 & 47.46 & 57.12 & 53.61 & 50.98 & 57.72 & 77.72 & 77.25 & 76.47 & 68.58 & 64.58 & 61.92 & 67.77 & 66.21 \\
    Ojha et al. \citeappx{Ojha_2023_CVPR} & 100.00 & 99.46 & 99.59 & 97.24 & 99.98 & 99.60 & 82.45 & 61.32 & 79.02 & 96.72 & 99.00 & 87.77 & 99.14 & 92.15 & 99.17 & 94.74 & 95.34 & 94.57 & 97.15 & 93.38 \\
    \rowcolor[gray]{0.9} Ours & 99.43 & 96.62 & 76.30 & 92.73 & 85.79 & 97.20 & 83.64 & 74.87 & 77.59 & 62.52 & 71.44 & 90.63 & 92.31 & 80.78 & 92.52 & 95.65 & 96.38 & 95.97 & 78.79 & 86.38 \\
    \bottomrule
    \end{tabular}
  }
   \caption{\textbf{Generalization Results (Average Precisions).} This table extends the analysis from \Cref{tab:accuracy_comparison}, focusing on the avearge precision of our detector when applied to diverse datasets after training on proGAN. Our method demonstrates superior precision compared to most alternatives and achieves results comparable to the state-of-the-art approach, underscoring its effectiveness in minimizing false positives across various image datasets.}
   
  \label{tab:precision_comparison}
   
\end{table*}